\documentclass[twocolumn,prl]{revtex4}
\usepackage{graphicx}
\usepackage{bm}
\usepackage {amsmath}
\usepackage {amssymb}

\newcommand* {\ket}[1]{\ensuremath{\left| {#1} \right\rangle}}
\newcommand* {\matrixelk}[3]{\ensuremath{\langle {#1} | {#2} | {#3} \rangle}}

\newcommand* {\vek}[1]{\ensuremath{\bm{\mathrm{#1}}}}

\newcommand* {\kk}{\vek{k}}
\newcommand* {\rr}{\vek{r}}
\newcommand* {\ee}{\ensuremath{\mathrm{e}}}
\newcommand* {\imag}{\ensuremath{i}}
\newcommand* {\bohrmag}{\mu_\mathrm{B}}
\newcommand* {\fermi}{\mathrm{F}}

\begin{document}

\title{Spin Polarization of Quasi Two-Dimensional Hole Systems}

\author{R.~Winkler}
\affiliation {Institut f\"{u}r Festk\"orperphysik,
Universit\"{a}t Hannover, Appelstr.~2, D-30167
Hannover, Germany}

\date{January 6, 2004}

\begin{abstract}
  In quasi two-dimensional (2D) hole systems with an effective spin
  $j=3/2$, heavy hole--light hole splitting results in a
  quantization of angular momentum perpendicular to the 2D plane.
  The spin polarization of quasi 2D hole systems due to an in-plane
  magnetic field $B_\|$ thus competes with the heavy hole--light
  hole splitting. As a result, the spin polarization of hole systems
  is very different from the spin polarization of $j=1/2$ electron
  systems.  In particular, it is shown that the spin polarization of
  quasi 2D heavy hole systems can change its sign at a finite value
  of $B_\|$.
\end{abstract}
\maketitle


A spin splitting of the energy levels of the electrons in a solid
can be caused by an external magnetic field $B$, the Zeeman
splitting, and we can also have a spin splitting due to the
inversion asymmetry (IA) of the crystal structure \cite{dre55a} or
the confining potential \cite{byc84}. The IA-induced spin splitting
can be ascribed to an effective magnetic field \cite{win03b}.
Intimately related with the spin splitting of the energy levels is
the spin orientation or spin polarization of the wave functions.
These quantities are of great importance for our understanding of
the fundamental properties of electron systems. Recently, the
interest in this subject has been renewed because it has been noted
that the spin polarization of quasi two-dimensional (2D) systems due
to a magnetic field $B_\|$ parallel to the 2D plane affects the
apparent metallic behavior of these systems \cite{kra04}.
Furthermore, the spin polarization is an important parameter for
possible applications in the field of spintronics \cite{wol01}. We
distinguish here between the \emph{spin orientation} which is a
property of the individual (occupied or empty) states, and the
\emph{spin polarization} which is the average spin orientation of
all occupied states. In quasi 2D systems the spin polarization has
been studied by applying a magnetic field $B_\|$ parallel to the 2D
plane \cite{smi72,oka99,tut02,pud02,zhu03,tut03,pap00a,tut01,pro02,%
noh03}. If the Zeeman energy splitting due to $B_\|$ becomes
sufficiently large, the minority spin subband is completely
depopulated. In such a situation, electron systems are fully
spin-polarized \cite{tut02,pud02,zhu03,tut03}.

Recently, there has been considerable interest in the spin
polarization of quasi 2D hole systems
\cite{pap00a,tut01,pro02,noh03}. Hole systems in the uppermost
valence band of many common semiconductors such as Ge and GaAs are
different from electron systems due to the fact that the hole states
have an effective spin $j=3/2$ \cite{lut56}. Subband quantization in
2D hole systems yields a quantization of angular momentum with $z$
component of angular momentum $m= \pm 3/2$ for the so-called heavy
hole (HH) states and $m= \pm 1/2$ for the light hole (LH) states.
Often it is assumed that the spin degree of freedom of HH and LH
states in quasi 2D hole systems behaves similar to the spin of $j =
1/2$ electron states. This assumption appears to be justified when
the HH-LH splitting is large. However, we will show here that such
an approach can be misleading. The important point is that the
quantization axis of angular momentum that is enforced by HH-LH
splitting is pointing perpendicular to the plane of the quasi 2D
system. On the other hand, in general the effective Hamiltonians for
IA-induced spin splitting and Zeeman splitting at $B_\| >0$ tend to
orient the spin vector parallel to the plane of the quasi 2D system
\cite{win03b}. The energy scale for IA-induced spin splitting and
Zeeman splitting are typically much smaller than the HH-LH
splitting. Thus it has been noted that for HH states the IA-induced
spin splitting \cite{win00a} and the Zeeman splitting at $B_\| > 0$
\cite{kes90} are suppressed, i.e., in a model with axial symmetry,
the IA-induced spin splitting and the Zeeman splitting are
higher-order effects proportional to $k_\|^3$ and $B_\|^3$,
respectively \cite{almost}. Here $k_\|$ is the in-plane wave vector.
This result reflects the fact that we can have only one quantization
axis of angular momentum.
A detailed discussion of HH-LH splitting can be found in
Ref.~\cite{win03a}.

In the present work we thus want to investigate the spin orientation
and spin polarization that can be achieved in quasi 2D hole systems.
We will show that in quasi 2D hole systems these quantities behave
fundamentally different from what is well-known about these
quantities in spin-$1/2$ electron systems. Throughout, we will
assume that only one orbital subband is occupied.

First, we briefly review for comparison the main features of spin
orientation and spin polarization in electron systems. In $j=1/2$
electron systems at $B=0$ we have two degenerate states
$\ket{\uparrow}$ and $\ket{\downarrow}$ that correspond to spin up
and spin down.  A finite magnetic field $\vek{B}$ gives rise to a
Zeeman term $\propto \vek{\sigma} \cdot \vek{B}$ in the Hamiltonian
which results in a spin splitting $\propto \pm B$ of the energy
levels. Here $B$ can be either an external magnetic field or the
$\kk$-dependent effective magnetic field due to a spin-orbit
interaction such as the Dresselhaus \cite{dre55a} or Rashba term
\cite{byc84}. The symbol $\vek{\sigma}$ denotes the vector of Pauli
spin matrices. The eigenstates $\ket{\pm}$ of the Hamiltonian at
$B>0$ are orthonormal linear combinations of the states
$\{\ket{\uparrow}, \ket{\downarrow}\}$ with spin orientation
$\vek{S}_+ \equiv \matrixelk{+}{\vek{\sigma}}{+}$ parallel to
$\vek{B}$ and $\vek{S}_- \equiv \matrixelk{-}{\vek{\sigma}}{-}$
antiparallel to $\vek{B}$. Furthermore, it is easy to show that for
normalized states $\ket{\pm}$ the spin orientations $\vek{S}_\pm$
are always unit vectors \cite{win03b}. We see here that the spin
orientation vectors $\vek{S}_\pm$ of electron states always follow
the magnetic field $\vek{B}$. The spin polarization at $B>0$ is
obtained by averaging over the spin orientation of all occupied
states. In the presence of an external magnetic field $\vek{B}_\|$
parallel to the 2D plane, the system will be fully spin polarized if
the Zeeman energy splitting is large enough to completely depopulate
the minority spin subband \cite{tut02}. If $\vek{B}_\|$ is an
effective magnetic field describing the spin splitting in
inversion-asymmetric systems, the $\kk_\|$ dependence of
$\vek{B}_\|$ is such that the spin polarization is zero
\cite{win03b}.

Hole states with an effective spin $j=3/2$ are characterized by the
$4\times 4$ Luttinger Hamiltonian \cite{lut56}. The corresponding
wave functions are four-component spinors. In quasi 2D systems we
get HH and LH states as discussed above. At $B=0$ the HH and LH
states are each two-fold degenerate. We can choose the eigenstates
\begin{subequations}
\label{eq:envelope_qw}
\begin{eqnarray}
\ket{\Psi_{\alpha\pm}^\mathrm{HH}} & = & 
\frac{\ee^{\imag \kk_\| \cdot \rr_\|}}{2 \pi} \;
\xi_{\alpha}^\mathrm{HH} (z) \; \frac{1}{\sqrt{2}}
\left(\begin{array}{c}
1 \\ 0 \\ 0 \\ \pm 1
  \end{array}  \right) \\
\ket{\Psi_{\alpha\pm}^\mathrm{LH}} & = & 
\frac{\ee^{\imag \kk_\| \cdot \rr_\|}}{2 \pi} \;
\xi_{\alpha}^\mathrm{LH} (z) \; \frac{1}{\sqrt{2}}
\left(\begin{array}{c}
 0 \\ 1 \\ \pm 1  \\ 0
  \end{array}  \right) .
\end{eqnarray}
\end{subequations}
Here we have assumed that the Luttinger Hamiltonian is expressed in
a basis of $j=3/2$ angular momentum eigenfunctions in the order
$m=3/2$, $1/2$, $-1/2$, and $-3/2$; and the quantization axis of
these basis functions is perpendicular to the 2D plane. In Eq.\ 
(\ref{eq:envelope_qw}) the symbol $\alpha$ denotes the subband
index. For simplicity we have neglected HH-LH mixing due to a
nonzero in-plane wave vector $\kk_\|$ and we also neglected the
$\kk_\|$ dependence of the envelope functions $\xi_{\alpha} (z)$.
These aspects are fully taken into account in the numerical
calculations discussed below. One can easily convince oneself that,
in contrast to $j=1/2$ electron systems, one can only find a linear
combination of the degenerate HH states
$\ket{\Psi_{\alpha+}^\mathrm{HH}}$ and
$\ket{\Psi_{\alpha+}^\mathrm{HH}}$ that are eigenstates of $J_z$,
but there are no such linear combinations that represent eigenstates
of $J_x$ or $J_y$. Here $J_x$, $J_y$, and $J_z$ are angular momentum
matrices for $j=3/2$. This fact clearly illustrates the frozen
angular momentum of HH states.

A finite in-plane magnetic field $B_\|$ gives rise to a mixing
of HH and LH states. In first order degenerate perturbation theory
we obtain the following expressions for the eigenstates of quasi 2D
holes in an in-plane field $B_\|$ in $x$ direction
\begin{subequations}
\label{eq:envelope_qw_b}
\begin{eqnarray}
\ket{\Psi_{\alpha\pm}^\mathrm{HH}} & = & 
\frac{\ee^{\imag \kk_\| \cdot \rr_\|}}{2 \pi} \;
\xi_{\alpha}^\mathrm{HH} (z) \; \frac{1}{\sqrt{2}}
\left(\begin{array}{c}
1 \\ - \sqrt{3} \mathcal{K} \mp \sqrt{3} \mathcal{G} \\ 
\mp \sqrt{3} \mathcal{K} - \sqrt{3} \mathcal{G} \\ \pm 1
  \end{array}  \right) \\
\ket{\Psi_{\alpha\pm}^\mathrm{LH}} & = & 
\frac{\ee^{\imag \kk_\| \cdot \rr_\|}}{2 \pi} \;
\xi_{\alpha}^\mathrm{LH} (z) \; \frac{1}{\sqrt{2}}
\left(\begin{array}{c}
 \sqrt{3}\mathcal{K} \pm \sqrt{3} \mathcal{G} \\ 1 \\ 
 \pm 1  \\ \pm \sqrt{3} \mathcal{K} + \sqrt{3} \mathcal{G}
  \end{array}  \right) ,
\end{eqnarray}
\end{subequations}
where
\begin{subequations}
  \label{eq:fak}
  \begin{eqnarray}
   \mathcal{K} & \equiv & 
   \frac{\kappa \bohrmag B_\|}{E^h_\alpha - E^l_\alpha}, \\
   \mathcal{G} & \equiv & \gamma_2
   \frac{2 m_0 \matrixelk{\xi_{\alpha}^\mathrm{HH}}{z^2}
                         {\xi_{\alpha}^\mathrm{LH}} }{\hbar^2} 
   \frac{(\bohrmag B_\|)^2}{E^h_\alpha - E^l_\alpha}
 \end{eqnarray}
\end{subequations}
are positive constants, $\kappa$ is the isotropic $g$ factor and
$\gamma_2$ is the second Luttinger parameter in the Luttinger
Hamiltonian \cite{lut56}, $\bohrmag$ is the Bohr magneton, $m_0$ is
the mass of free electrons, $E^h_\alpha$ ($E^l_\alpha$) is the
energy of the $\alpha$th HH (LH) subband at $B_\|=0$, and we have
used the gauge $\vek{A} = (0 , -z B_\|, 0)$. In our notation, the HH
``+'' states ($\alpha=1$) belong to the majority spin subband that
remains occupied for large $B_\|$. For simplicity, we have neglected
the small terms proportional to the anisotropic $g$ factor $q$. We
also neglected the coupling to neighboring subbands $\alpha' \ne
\alpha$. We note that $\mathcal{K} \propto w^2$ and $\mathcal{G}
\propto w^4$, where $w$ is the width of the quasi-2D layer.

Similar to electron states the spin orientation of hole states is
given by \cite{win03b}
\begin{equation}
  \label{eq:spin_hole}
  \vek{S}_{\alpha\pm} = \frac{2}{3}
\matrixelk{\Psi_{\alpha\pm}}{\vek{J}} {\Psi_{\alpha\pm}} \; .
\end{equation}
Here $\vek{J} = (J_x, J_y, J_z)$ is the vector of angular momentum
matrices for $j=3/2$ and the prefactor $2/3$ ensures that
$|\vek{S}_{\alpha\pm}| \le 1$. For a magnetic field in $x$ direction
only the $x$ components of the vectors $\vek{S}_{\alpha\pm}$ are
nonzero. (As discussed below we also get a $y$ component if the
$\kk_\|$ dependence of $\ket{\Psi_{\alpha\pm}}$ is taken into
account.) Using Eq.\ (\ref{eq:envelope_qw_b}) we obtain in first
order of the HH-LH splitting $E^h_\alpha - E^l_\alpha$
\begin{subequations}
  \label{eq:hole_orient}
  \begin{eqnarray}
  \label{eq:hh_orient}
    S^\mathrm{HH}_{\alpha\pm} & = & - 2 \mathcal{K} \mp 2 \mathcal{G}  \\
  \label{eq:lh_orient}
    S^\mathrm{LH}_{\alpha\pm} & = & 
    \pm 2/3 + 2 \mathcal{K} \pm 2 \mathcal{G} \; .
  \end{eqnarray}
\end{subequations}
For typical values of $B_\|$ the second and third term in Eq.\ 
(\ref{eq:lh_orient}) are much smaller than $2/3$. Thus, similar to
electron states at $B_\| > 0$, the spin orientations
$S^\mathrm{LH}_{\alpha +}$ and $S^\mathrm{LH}_{\alpha -}$ of the
Zeeman-split LH states are essentially antiparallel to each other
and independent of $B_\|$. For small fields $B_\|$ the Zeeman-split
HH states have the \emph{same} spin orientation antiparallel to
$\vek{B}_\|$ which is ``compensated'' by a term of opposite sign in
Eq.\ (\ref{eq:lh_orient}). The spin orientations
$S^\mathrm{HH}_{\alpha\pm}$ vanish for $B_\| \rightarrow 0$. We see
here that HH states are fundamentally different from $j=1/2$
electron states where the spin orientation vectors of the spin-split
states are always antiparallel to each other, and they have always
length one, independent of the magnitude of $B_\|$ \cite{win03b}.
According to Eq.\ (\ref{eq:hh_orient}) the spin orientation vectors
of HH states are the smaller in magnitude the larger the HH-LH
splitting $E^h_\alpha - E^l_\alpha$, which is opposite to what one
expects according to the picture that the spin of hole systems with
a large HH-LH splitting behaved similar to the spin of $j = 1/2$
systems. Once again, these results reflect the frozen angular
momentum of hole states due to the HH-LH splitting. For larger
fields $B_\|$ the term $\mathcal{G}$ quadratic in $B_\|$ dominates
over the linear term $\mathcal{K}$ in Eq.\ (\ref{eq:hh_orient}). For
the HH ``+'' states, we thus obtain a sign reversal of the spin
orientation $S^\mathrm{HH}_{\alpha +}$ at a finite value of $B_\|$.
The term $\mathcal{K}$ is proportional to the hole $g$ factor
$\kappa$ whereas $\mathcal{G}$ is proportional to the Luttinger
parameter $\gamma_2$ characterizing the orbital motion of the holes.
We can thus attribute the sign reversal of $S^\mathrm{HH}_{\alpha
+}$ to a competition between the effect of the motion of the orbital
and the spin degree of freedom. We remark here that usually only the
first HH subband is occupied in 2D hole systems. The sign reversal
of $S^\mathrm{HH}_{\alpha +}$ as a function of $B_\|$ is thus an
important feature of these systems. It depends sensitively on the
width $w$ of the quasi-2D layer.

The functional form of the Zeeman energy splitting in an in-plane
magnetic field $B_\|$ is very different from the expressions for the
spin orientation in Eq.\ (\ref{eq:hole_orient}). We emphasize here
that the spin orientation depends on the eigenfunctions of the
Schr\"odinger equation whereas the Zeeman splitting depends
on the eigenvalues. In second order perturbation theory the Zeeman
energy splitting due to an in-plane field $B_\|$ in $x$ direction
reads~\cite{win03a}
\begin{subequations}
  \label{eq:zeeman}
  \begin{eqnarray}
    \Delta E^\mathrm{HH}_{\alpha} & = & - \mathcal{Z} \\
    \Delta E^\mathrm{LH}_{\alpha} & = &
   - 4 \kappa \bohrmag B_\| + \mathcal{Z} ,
  \end{eqnarray}
\end{subequations}
where
\begin{equation}
  \label{eq:zeemanfak}
  \mathcal{Z} \equiv \kappa \gamma_2
     \frac{24 m_0 \matrixelk{\xi_{\alpha}^\mathrm{HH}}{z^2}
                            {\xi_{\alpha}^\mathrm{LH}}}{\hbar^2}
     \frac{(\bohrmag B_\|)^3}{E^h_\alpha - E^l_\alpha}
\end{equation}
The product of $\kappa$ and $\gamma_2$ in Eq.\ (\ref{eq:zeemanfak})
as well as the cubic dependence on $B_\|$ indicate that in contrast
to the spin orientation of HH states the Zeeman energy splitting is
due to the \emph{combined} effect of the motion of the orbital and
the spin degree of freedom. The expressions for the energy
eigenvalues also contain terms of even order in $B_\|$. However,
these terms do not contribute to the Zeeman splitting. We note that
$\mathcal{Z} \propto w^4$.

If the spin splitting is caused by the $\kk_\|$-dependent effective
magnetic field in inversion-asymmetric systems we obtain results
similar to Eqs.\ (\ref{eq:hole_orient}) and (\ref{eq:zeeman}) with
$B_\|$ replaced by $k_\|$. In particular, once again the spin
vectors $\vek{S}_{\alpha\pm}^\mathrm{HH} (\kk_\|)$ of the spin-split
HH states for a given in-plane wave vector $\kk_\|$ are in general
neither unit vectors nor they are antiparallel to each other.

We remark that previously it has proven useful to discuss spin
splitting in quasi 2D HH systems by means of $2\times 2$
Hamiltonians acting only in the subspace of HH states \cite{win03a}.
On the other hand, in the present discussion it is crucial that we
use the full $4 \times 4$ Luttinger Hamiltonian to evaluate the spin
orientation of quasi 2D systems. Here it should be kept in mind that
the $2\times 2$ Hamiltonians are \emph{effective} Hamiltonians that
emerge from a perturbative unitary transformation of the full $4
\times 4$ Luttinger Hamiltonian in order to decouple the motion of
HH and LH states. To evaluate the spin orientation $\vek{S}$ we must
either express $\vek{J}$ in the transformed basis or we must
(re-)express the HH wave functions as four-component spinors as in
Eqs.\ (\ref{eq:envelope_qw}) and (\ref{eq:envelope_qw_b}).

\begin{figure}[t]
  \includegraphics[width=0.77\columnwidth]{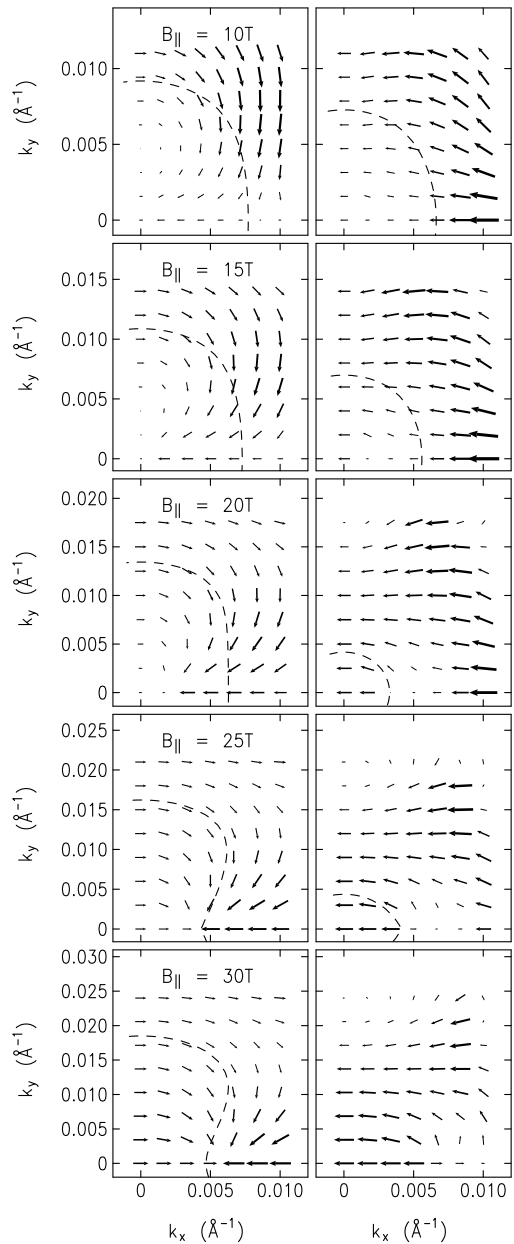}
  \caption{\label{fig:hh_ori} Spin orientation
  $\vek{S}_{1\pm}^\mathrm{HH}$ as a function of $\kk_\|$ for a
  symmetric (100) GaAs-Al$_{0.3}$Ga$_{0.7}$As quantum well with hole
  density $p = 1 \times 10^{11}$~cm$^{-2}$ and different values of
  an in-plane magnetic field $B_\|$ in $x$ direction. The left
  (right) column refers to the majority (minority) spin subband. The
  dimensions of the arrows are proportional to
  $\vek{S}_{1\pm}^\mathrm{HH} (\kk_\|)$. The dashed lines indicate
  the $B_\|$-dependent Fermi contours.}
\end{figure}

To confirm the qualitative results in Eq.\ (\ref{eq:hole_orient}) we
show in Fig.~\ref{fig:hh_ori} the self-consistently calculated spin
orientation $\vek{S}_{1\pm}^\mathrm{HH}$ as a function of $\kk_\|$ for a
symmetric (100) GaAs-Al$_{0.3}$Ga$_{0.7}$As quantum well with hole
density $p = 1 \times 10^{11}$~cm$^{-2}$ and different values of an
in-plane magnetic field $B_\|$ in $x$ direction.
The left (right) column refers to the majority (minority) spin
subband. The dashed lines show the Fermi contours. The numerical
calculations follow Ref.\ \cite{win03a}. They fully take into
account the anisotropy of the crystal structure. We see in
Fig.~\ref{fig:hh_ori} that the magnitude and the orientation of
$\vek{S}_{1\pm}^\mathrm{HH} (\kk_\|)$ depend sensitively on $B_\|$
and $\kk_\|$, in sharp contrast to electron states. For small values
of $B_\|$ and $k_\|$, the spin orientation vectors of the HH states
in both spin subbands are essentially antiparallel to $\vek{B}_\|$,
in agreement with Eq.\ (\ref{eq:hh_orient}). For larger values of
$k_\|$ we can have a finite angle between
$\vek{S}_{1\pm}^\mathrm{HH} (\kk_\|)$ and $\vek{B}_\|$. Furthermore,
the spin vectors $\vek{S}_{1+}^\mathrm{HH} (\kk_\|)$ and
$\vek{S}_{1-}^\mathrm{HH} (\kk_\|)$ for a given wave vector $\kk_\|$
are in general neither parallel nor antiparallel. We remark that the
$\kk_\|$ dependence of the spin orientation can be studied
analytically using the model underlying Eqs.\ 
(\ref{eq:envelope_qw_b}) and (\ref{eq:hole_orient}). However, we do
not reproduce here the lengthy formulas.

\begin{figure}[t]
  \includegraphics[width=0.7\columnwidth]{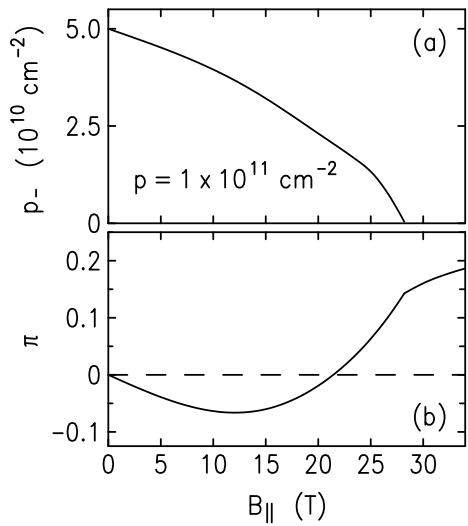}
  \caption{\label{fig:hh_pol}(a) Spin subband density
  $p_-$ of the HH minority spin subband and (b) spin polarization
  $\pi$ as a function of the in-plane field $B_\|$ calculated for a
  symmetric (100) GaAs-Al$_{0.3}$Ga$_{0.7}$As quantum well with hole
  density $p = 1 \times 10^{11}$~cm$^{-2}$.}
\end{figure}

For the same system as in Fig.~\ref{fig:hh_ori} we show in
Fig.~\ref{fig:hh_pol}(a) the self-consistently calculated
depopulation of the HH minority spin subband,
and Fig.~\ref{fig:hh_pol}(b) shows the $x$ component of the
normalized spin polarization
\begin{equation}
  \label{eq:pol}
  \vek{\pi} = \frac{1}{p}
  \sum_{\alpha,\pm}
  \int \! d k_\|^2 \:
  \theta [E_{\alpha\pm} (\kk_\|) - E_\fermi] \;
  \vek{S}_{\alpha\pm} (\kk_\|) \; .
\end{equation}
Here $E_{\alpha\pm} (\kk_\|)$ is the spin-split subband dispersion
and $E_\fermi$ is the Fermi energy. If $\vek{B}_\|$ is pointing in
$x$ direction, the $y$ and $z$ components of $\vek{\pi}$ are zero.
In Eq.\ (\ref{eq:pol}) we have assumed that hole energies are
negative and the temperature is zero. We evaluate Eq.\ 
(\ref{eq:pol}) by means of analytic quadratic Brillouin zone
integration \cite{win93}. It becomes clear from
Fig.~\ref{fig:hh_pol} that the depopulation of the HH minority spin
subband due to $B_\|$ does not imply full spin polarization of the
system.  In Fig.~\ref{fig:hh_pol} the minority spin subband is
completely depopulated at $B_D \approx 28.2$~T while $\pi (B_D)
\approx 0.14$. In agreement with Eq.\ (\ref{eq:hh_orient}) we obtain
a sign reversal of $\pi$ at $B_\| \approx 21.4$~T. The derivative of
$\pi (B_\|)$ is discontinuous at $B_\| = B_D$.

We have checked that the surprising results reported here are
observed for a large variety of 2D HH systems, although the details
depend on system parameters such as the density $p$, the width and
the crystallographic growth direction of the quasi 2D system, and
the material composition of the layers. In the calculations, we have
neglected many-particle effects which will be important for a
quantitative analysis of spin polarization in the density regime
discussed here. However, to the best of our knowledge an appropriate
theory for particles with $j=3/2$ is presently not available. We
expect that the essential features in Figs.~\ref{fig:hh_ori}
and~\ref{fig:hh_pol} will be confirmed by such a more quantitative
theory.

The author appreciates stimulating discussions with E.~Tutuc and
M.~Shayegan. This work was supported by BMBF.


\begin{thebibliography}{21}
\bibitem[{alm()}]{almost}
\bibinfo{note}{We get a splitting of HH states linear in $k_\|$ or $B_\|$ if
  the crystal anisotropy is taken into account. This reflects the fact that
  angular momentum in Eq.\ (\ref{eq:envelope_qw}) is merely an almost good
  quantum number~\cite{win03a}.}
\end{thebibliography}
\end{document}